\documentclass[12pt,english]{scrartcl}

\usepackage[OT1]{fontenc}
\usepackage[latin1]{inputenc}
\usepackage{amsmath}
\usepackage{graphicx}
\usepackage{babel}
\usepackage{amssymb}
\IfFileExists{url.sty}{\usepackage{url}}
                      {\newcommand{\url}{\texttt}}

\makeatletter

\newif \ifpdf
\ifx \pdfoutput \undefined
      \pdffalse
   \else
      \pdftrue
\fi
\ifpdf
\usepackage{mathptmx}
\usepackage{helvet}
\usepackage{ae,aecompl}
\usepackage[pdftex,urlcolor=blue,colorlinks=true]{hyperref}
\else
\usepackage[ps2pdf,urlcolor=blue,colorlinks=true]{hyperref}
\fi

\title{Spatial and polarization structure in micro-dome resonators: effects of a Bragg mirror} 

\author{David H. Foster and Jens U. N\"{o}ckel\\
\medskip
Oregon Center for Optics,\\
1274 University of Oregon,\\ 
Eugene, OR 97403-1274\\
\url{http://darkwing.uoregon.edu/~noeckel}
}
\date{Published in {\em Laser Resonators and Beam Contol VII}, edited
  by Alexis V. Kudryashov and Alan H. Paxton, Proceedings of SPIE {\bf
    5333}, 195-203 (2004)}

\begin{document} 
  \maketitle 

\begin{abstract}
Micro-domes based on a combination of metallic and dielectric
multilayer mirrors are studied using a fully vectorial numerical
basis-expansion method that accurately accounts for the effects of an
arbitrary Bragg stack and can efficiently cover a large range of dome
shapes and sizes. Results are examined from three different
viewpoints: (i) the ray-optics limit, (ii) the (semi-) confocal limit
for which exact wave solutions are known, and (iii) the paraxial
approximation using vectorial Gaussian beams. 
\end{abstract}


\section{Introduction}

Microresonators based on planar distributed Bragg reflectors (DBR) are ubiquitous
because they afford great design flexibility, e.g., tailored stopbands
of ultrahigh reflectivity, and at the same their fabrication is well-developed
in many material systems. In many applications, however, control over
transverse mode profiles and sidemodes is improved when non-planar
structures are used. Stable, dome-shaped cavities have been employed
with InGaAs quantum-well lasers \cite{Matinaga}, and as passive filter
cavites \cite{Meissner}. Such cavities are expected to be of great
promise for semiconductor-based quantum optics \cite{milburn}, because
strong focusing and large modulation of the local density of states
can be achieved \cite{Jens}. Because dome shaped DBR structures are
technologically challenging, combinations of flat DBR stacks and curved
metal mirrors have been considered as a compromise solution\cite{Jens}.
Planar metal layers have recently proven useful as top mirrors in
thin-film organic lasers \cite{bulovic}, and novel fabrication schemes
\cite{coyle} are likely to make curved metal mirrors a building block
in microcavity design. In this paper, we address dome cavities in
which metal mirrors and Bragg stacks are combined. 

Computationally, such cavities are nontrivial even when the medium
is linear, because they combine an unconventional three-dimensional
shape with boundary conditions that differ at the top and bottom of
the structure. Realistic Bragg mirrors of finite depth may also permit
significant leakage, which has to be modeled accurately in order to
determine the coupling properties of the resonator. We have performed
fully vectorial electromagnetic simulations of axially symmetric cavities
formed by a curved metallic dome on a stratified Bragg stack. The
composition of the stack mirror and the shape of the metal dome have
been varied in a wide range of parameters, and the capabilities of
the numerical methods have been tested up to cavity lengths of $50\lambda$,
where $\lambda$ is the wavelength in the dielectric comprising the
bulk of the dome. 

\section{Cavity geometry}

\begin{figure}[!htb]
     \begin{center}
 
\includegraphics[%
  width=0.4\columnwidth]{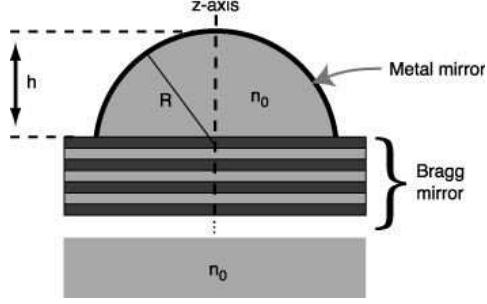}
  
   \end{center}

\caption{\label{cap:Geometry}Geometry of the dome resonator with a
  Bragg mirror at its base. The dome mirror is taken to be a spherical
  shell of radius $R$, centered at a depth $R-h$ below the planar 
  mirror surface. The bulk of the dome, and the substrate have
  refractive index $n_0$.}
\end{figure}

The cavity geometry is shown in Fig. \ref{cap:Geometry}. It is useful
to define a cylindrical coordinate system ($\rho$, $\phi$, $z$)
with $z$ being the symmetry axis. The height $h$ of the dome above
the Bragg mirror, and its (not necessecarily constant) radius of curvature,
$R$, determine the type of modes the cavity can support. For definiteness,
in this work the bulk of the dome and the space outside of the structure
have refractive index $n_{0}=1$. In the numerical results shown below,
we will adjust the the Bragg stack design such that its stop band
is centered at the modes of interest. The DBR layer structure begins
with a spacer layer of optical thickness $\lambda_{s}$ and refractive 
index $3.5$, followed
by $N_{s}=20$ quarter-wave layer pairs $AB$ of
refractive index $n_{A}=3.0$, $n_{B}=3.5$. Our main goal is
to demonstrate the importance of the polarization-dependent phase
shifts induced by this mirror in determining the resonator modes.
Therefore, we also performed comparative calculations with the Bragg
stack replaced by a conducting surface. In accordance with this main
goal of the present study, absorption in the metal surfaces was neglected.
Below, the curved mirror will be taken to be a spherical shell of
constant $R$. However, we have also carried out calculations for
domes in the shape of a rotational paraboloid, to compare with earlier
calculations based on a separation of variables that is possible for
that system \cite{Jens} when only scalar fields are considered. At
size parameters between $kR=26\ldots31$, and using only metal mirrors,
all modes agreed in their spectral positions to better than five decimal
places, despite the very different computational approach used in
the present work.

\section{Vector fields in the dome resonator}

We are looking for the divergence-free, monochromatic ($\propto\exp[-\imath\omega t]$)
electric fields satisfying the vectorial wave equation. Using the
axial symmetry, the field components $E_{\rho}$, $E_{\phi}$ and
$E_{z}$ can be assumed to have a $\phi$-dependence of the form $\exp(im\phi)$.
For fixed azimuthal mode number $m$, the electric field \emph{transverse}
to the $z$ axis, $\mathbf{E}_{T}=E_{\rho}\hat{\rho}+E_{\phi}\hat{\phi}$,
can be decomposed into circularly polarized components using \begin{eqnarray*}
E_{\rho} & = & \frac{\imath}{\sqrt{2}}(E_{+}-E_{-})\\
E_{\phi} & = & \frac{1}{\sqrt{2}}(E_{+}+E_{-})\end{eqnarray*}
where \begin{equation}
\rho^{2}\left[\nabla^{2}+k^{2}\right]E_{\pm}=(1\pm2m)E_{\pm}\label{eq:circwave}\end{equation}
With this, one has \begin{equation}
\mathbf{E}_{T}=\frac{1}{2}\underbrace{(E_{\rho}+\imath E_{\phi})}_{\propto\,\exp(\imath m\phi)}e^{\imath\phi}\left(\begin{array}{c}
1\\
-\imath\end{array}\right)+\frac{1}{2}\underbrace{(E_{\rho}-\imath E_{\phi})}_{\propto\,\exp(\imath m\phi)}e^{-\imath\phi}\left(\begin{array}{c}
1\\
\imath\end{array}\right)\label{eq:transversesphamrexp}\end{equation}
The standard conducting-mirror boundary conditions on the curved dome
($E_{\parallel}=\mathbf{0}$, $H_{\perp}=\mathbf{0}$),
together with the planar mirror at the base of the dome, generally
couple $E_{\rho}$, $E_{\phi}$ and $E_{z}$. All modes can be labeled
by $m$, and the substitution $m\to-m$ (for $m\ne0$) leads to a
degenerate mode. In presenting results later on, there is no loss of
generality if $m\ge 0$ is assumed. 

As was found in \cite{Jens}, focusing on the $z$
axis is most pronounced for the states of the cavity in which either 
the magnetic or the electric 
field is polarized exlusively in the azimuthal direction $\hat{\phi}$.
If $\mathbf{B}\propto\hat{\phi}$, then $\mathbf{E}$ has components
only along $\hat{z}$ and $\hat{\rho}$.

\section{All-metal cavity}
\label{sec:All-metal-cavity}
\begin{figure}[!htb]

   \begin{center}
 
\includegraphics[%
  width=0.9\columnwidth]{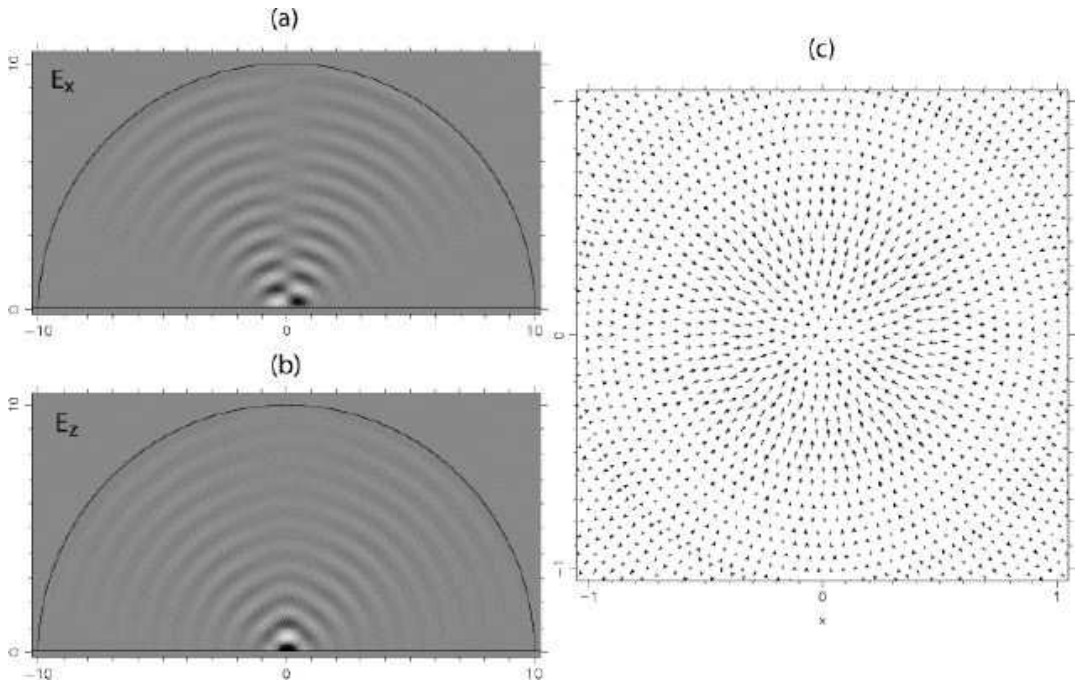}
  
   \end{center}
\caption{\label{cap:RadialMode}Grayscale plots of the $\mathbf{E}$-field
components (a) $E_{x}$ and (b) $E_{z}$ for a {}``radially polarized''
mode at $\lambda=867$nm in a conducting dome with $h=9.95\,\mu\mathrm{m}$,
$R=10\,\mu\mathrm{m}$. Tight focusing is apparent, especially in
$E_{z}$. Here and in the following vertical cross-sections, the field
is plotted in the $xz$ plane. The projection of the electric polarization
in a transverse ($xy$) plane at height $z=0.25\,\mu\mathrm{m}$ above the
planar metal mirror are shown in (c). }

\end{figure}

In a recent free-space beam experiment \cite{Dorn}, extremely strong
focusing was indeed observed with light having this type of polarization;
the beam was called ``radially polarized'', but a significant
contribution to the focused spot actually comes from $E_{z}$. The analogue
of this for our dome cavity is illustrated in Fig. \ref{cap:RadialMode},
assuming the flat mirror to be a conductor \footnote{In all wave plots
  that are shown,  we set $\omega t=\pi/2$ or $0$ and take the real parts of 
the electric fields. }. In the hemispherical limit, this mode evolves
smoothly into an even more strongly focused electric multipole 
with $m=0$ and $l=1$. 

In the generic cavity with $h<R$, any given mode will have admixtures
of many multipoles  with different $l$, both of electric and magnetic
type. The azimuthal mode number $m$ is a
fixed parameter in the calculation. To obtain Fig. \ref{cap:RadialMode},
the coefficients $\mathbf{y}$ of this basis expansion are determined
from a linear system of equations set up by the boundary conditions.
The number of unknowns is determined by the number of different $l=l_{min}\ldots l_{max}$in
the expansion. Here, $l_{min}=\mathrm{max}(1,|m|)$, and an approximate
cutoff $l_{max}$ is given by the semiclassical limit $kn_{0}\rho_{max}$
where $\rho_{max}$ is the maximum radial extent of the dome. Since
we are interested in cavity modes, there is no incoming wave outside
the dome, and the resulting system of equations is initially homogeneous.
Our method proceeds by adding a further {}``seed'' equation that
sets a certain linear combinations of unknowns equal to one \cite{thesis},
thus converting the problem into an inhomogeneous matrix equation
$A\mathbf{y}=\mathbf{b}$ where the matrix $A$ depends on $k$, and
$\mathbf{b}\neq\mathbf{0}$ because of the seed condition. The boundary
conditions are enforced by point matching (real-space points on the
curved mirror, and $\mathbf{k}$-space points for the planar mirror), 
and the wavenumbers of
the cavity modes are found by minimizing the residual $\sqrt{\vert A\mathbf{y}-\mathbf{b}\vert^{2}}$
as a function of $k$. Details of the numerical method are presented
in \cite{David}. Before discussing the additonal complications posed
by a dielectric stack mirror, we use the all-metal cavity to further
investigate the effect of the dome-shaped mirror.

\section{Relation to ray dynamics and confocal limit}

A global picture of the possible cavity modes can be obtained from
geometric optics. Mirror configurations with $h<R$ are called stable
because they support modes centered on stable axial rays, while $h=R$
creates a confocal geometry if one unfolds the cavity about the planar
mirror \cite{Jens} (one might alternatively refer to $h=R$ as the 
``semi-confocal'' condition, but ray-optically there is no difference
to a symmetric, confocal mirror arrangement). 
This marginally stable limiting case features
highly focused but non-paraxial modes. Non-paraxial ray orbits that
can support cavity modes are also found for $h<R$, and an efficient
tool to reveal all the coexisting types of stable mode patterns is
the Poincar{\'e} surface of section\cite{Jens}. 

The axial symmetry of
the cavity implies that any given ray trajectory $\mathbf{r}(s)$,
where $s$ is the path length, can be parametrized by an equation
of the form $\mathbf{r}\times d\mathbf{r}/ds=\mathbf{L}$ with a vector
$\mathbf{L}$ that changes upon reflection at the boundaries but whose
component $L_{z}$ is a constant of the motion (the skewness of the
rays). Thus $L_{z}$labels families of rays, and fundamental Gaussian-beam
type modes are built on rays with $L_{z}\approx0$. However, even
when $L_{z}$ is fixed, stable periodic ray orbits with a multitude
of different topologies will generally be found. 

\begin{figure}[!htb]
   \begin{center}
 
\includegraphics[%
  width=1.0\columnwidth]{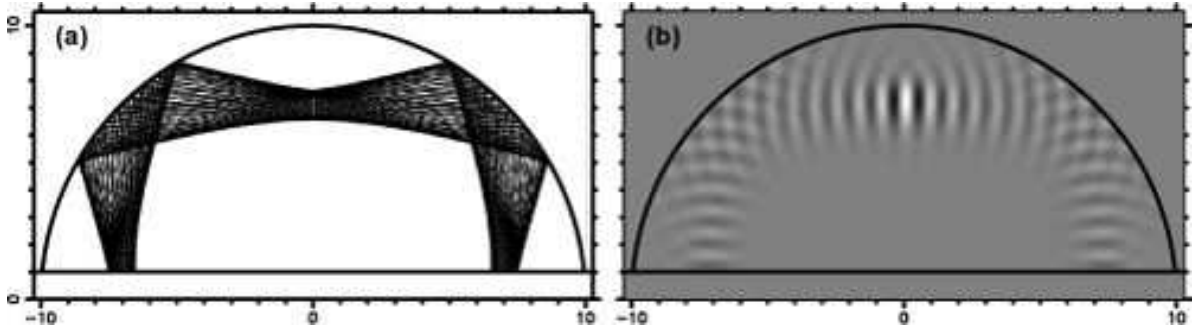}
\end{center}
\caption{\label{RectangleMode}(a) ray pattern (b) y-component of the electric
field (grayscale), plotted in a meridional cross section along the
xz plane. Axis labels are in microns. 
}
\end{figure}

Regarding the internal
ray dynamics, the dome cavity is thus closely related to axially symmetric
spheroidal resonators, such as droplets and microspheres \cite{mekis},
for which the confocal limit corresponds to perfect sphericity. This
analogy extends to the occurence of whispering-gallery type patterns
and their non-perturbative breakup under shape distortion, which has
recenty been observed in fused-silica microspheres \cite{lacey}.
As an illustration of this phenomenology in a dome cavity,
Fig. \ref{RectangleMode} 
shows a cavity with $h=9\,\mathrm{\mu m}$, $R=10\,\mu\mathrm{m}$.
The rectangular ray pattern in (a) for $L_{z}=0$ can be thought of
as a stabilized periodic whispering-gallery trajectory. 

A corresponding solution of the vectorial wave equation is shown in
Fig. \ref{RectangleMode} (b), indicating that the potential for strongly
focused modes in near-confocal cavities comes at the price of an increased
variety of off-axis modes which need to be characterized in order
to choose the optimal design. It is conceivable that such new mode
patterns are in fact desirable for certain applications. In order to
take advantage of this new variety of modes, it will be important to
combine 3D cavity shape design with suitably tailored Bragg
mirrors that discriminate against the unwanted type of modes. 
For paraboloidal domes, some initial exploration of this
aspect has been performed\cite{Jens}, in particular applying ray-optic
phase space analysis to the problem of angle-dependent Bragg mirror 
transmission. 

\section{Relation to the paraxial approximation}

However in the remainder of this paper, we turn our attention to the more
conventional 
modes centered on the $z$ axis. These can be understood in terms
of Gaussian beams when the paraxial parameter $p\equiv\lambda/z_{R}\ll1$,
where $z_{R}=\pi w_{0}$ is the the Rayleigh range and $w_{0}$ is
the beam waist radius \cite{Siegman}. An expression for the transverse
field at a fixed $m$ in terms of the vector Laguerre-Gaussian basis
can be written in analogy to (\ref{eq:transversesphamrexp}), 
\begin{equation}
\mathbf{E}_{T}=A\left(\begin{array}{c}
1\\
-\imath\end{array}\right)LG_{N}^{m+1}(\mathbf{r})+B\left(\begin{array}{c}
1\\
\imath\end{array}\right)LG_{N}^{m-1}(\mathbf{r})\label{eq:veclaguerregauss}
\end{equation}
Here, $A$ and $B$ are amplitude coefficients, the scalar Laguerre-Gauss
functions $LG_{N}^{m\pm 1}$ have a $\phi$ dependence
$\exp(\imath (m\pm 1)\phi)$, and  $N$ is the order of the Laguerre-Gauss
beam, 
$N=|m|+1,\,|m|+3,\,\ldots$. This definiton is related to a convenient
normalized expression $u_{k,l}^{LG}$ available in the literature
\cite{Beijersbergen} by $LG_{N}^{\mu}\equiv
u_{(N-\mu)/2,(N+\mu)/2}^{LG}$ (we have $\mu =m\pm 1$ for fixed
$m$). The special choice $N=|m|-1$ is also possible, provided $m\ne 0$; in
this case, admissible solutions must have $A=0$ if $m>0$, or $B=0$ if
$m<0$, so that $\mathbf{E}_{T}$ has circular polarization. 
As a consequence, free-space LG solutions {\em at given} $m$ with 
order $N=|m|-1$ are nondegenerate. For all other orders $N$ (and any $m$),
the two terms in (\ref{eq:veclaguerregauss})
are degenerate \emph{free-space} solutions in the paraxial approximation,
and will remain nearly degenerate in a paraxial \emph{cavity}. The
polarization properties of the mode are fixed by the 
way in which the boundary conditions split this doublet (for $m\ne
0$).

\begin{figure}[!htb]

   \begin{center}
 
\includegraphics[%
  width=0.65\columnwidth]{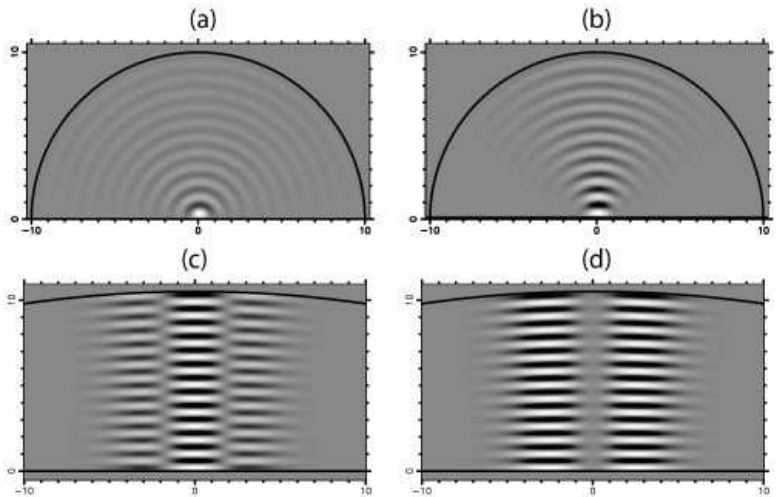}
  
   \end{center}
\caption{\label{cap:AllMetal}Some modes of the all-metal cavity with azimuthal
order $m=1$: fixing $R=10\,\mu\mathrm{m}$, a mode of type $\mathrm{TEM}_{00}$
narrows as $h$ decreases from (a) $h=9.9992$ ($\lambda=889.02$
nm) to (b) $h=9.9000$ ($\lambda=881.30$ nm). Compare to the analogous
results with a Bragg stack in Fig. \ref{cap:BraggSide}. Shown in
(c) is a paraxial mode of type $LG_{2}^{0}$ at $h=10.5\,\mu\mathrm{m}$,
$R=70\,\mu\mathrm{m}$ and $\lambda=796.060$nm. (d) shows its near-degenerate
partner $LG_{2}^{2}$ at $\lambda=796.055$nm. See Fig. \ref{cap:TransverseBragg}
for the transverse polarization patterns corresponding to (c) and
(d).}

\end{figure}

The fundamental Gaussian modes of the dome can be obtained by setting
$m=1,\, N=0$ and $A=0$. The spatial dependence
is then $E_{T}\propto LG_{0}^{0}$, which is also called a $\mathrm{TEM}_{00}$
mode. Figures \ref{cap:AllMetal} (a) and (b) show how the resemblance
to this Gaussian mode emerges starting from the hemispherical limit. 

Raising $N$ in increments of $2$ makes higher transverse excitations
of the Gaussian mode accessible. This is again borne out by the numerical
vectorial solutions, as shown for $m=1$ in Fig. \ref{cap:AllMetal}
(c) and (d). These modes are well described by (\ref{eq:veclaguerregauss})
with $N=|m|+1=2$ and either $A=0$ (c) or $B=0$ (d). Note that the
all-metallic boundary conditions have uniquely fixed the polarization
of the modes in Fig. \ref{cap:AllMetal} to be circular, i.e. only
one of the terms in (\ref{eq:veclaguerregauss}) is present in all
cases shown up to this point, and the degeneracy in $\lambda$ is
lifted by a small but finite amount. 

\section{Combining metal dome and Bragg mirror}

The previous results raise the question if all cavity modes of the
axially symmetric dome necessarily factor into a unique polarization
vector multiplied by a spatial wave function, thus making the wave
problem essentially scalar. We will demonstrate below that this is
\emph{not} the case when a Bragg stack replaces the planar metal mirror
at the base of the dome. To this end we first discuss the necessary
extension of the numerical method based on the basis expansion of
Section \ref{sec:All-metal-cavity}, allowing us to treat the combined
problem of metal dome and dielectric mirror without approximations. 

The complex reflection amplitude of an arbitrary planar Bragg grating
is calculated with the transfer matrix method \cite{Yeh} and can
be written in the plane-wave basis as a function of polarization,
wavelength and angle of incidence, $r_{s,p}(k,\theta)$. Here, subscripts
$s$ and $p$ distinguish between polarization perpendicular and in
the plane of incidence. Because of this simplicity, one may decide
to solve the whole dome problem in a discretized plane wave basis,
which after symmetrization according to the axial rotation invariance
leads to expansions of all vector field components in Bessel beams.
Their (scalar) form in the dome region is 
\begin{equation}
\psi_{Bessel}(\rho,\phi,z;\theta)=2\pi\imath^{m}\exp(\imath
m\phi)\,\exp(\imath n_{0}kz\cos\theta)\, J_{m}(\rho
n_{0}k\,\sin\theta),
\end{equation}
where $\theta$ is a parameter that specifies the cone of plane waves
from which this beam is constructed. 

However, the approach we follow in this paper is to \emph{combine}
Bessel waves and multipoles. We use the $\psi_{Bessel}$
to describe the DBR mirror, because reflection amplitudes for a Bessel
beam of given $\theta$ are identical to those of a plane wave at
that same incident angle, $r_{s,p}(\theta)$. But we retain multipoles
as the basis in which to specify the dome boundary
conditions, because these basis functions evolve into the true cavity
modes in the limiting case of the all-metal hemisphere. We discuss
elsewhere \cite{David} the relative merits of this approach compared
to the procedurally more direct method of using a single basis for the
entire domain. Our two-basis method makes it necessary
to (i) discretize the cone angles $\theta$ used in the expansions,
and (ii) to implement a transformation between multipoles 
and plane waves so that we effectively obtain a formulation of the
Bragg reflection in the multipole basis. 

\begin{figure}[!htb]
   \begin{center}
 
\includegraphics[%
  width=0.65\columnwidth]{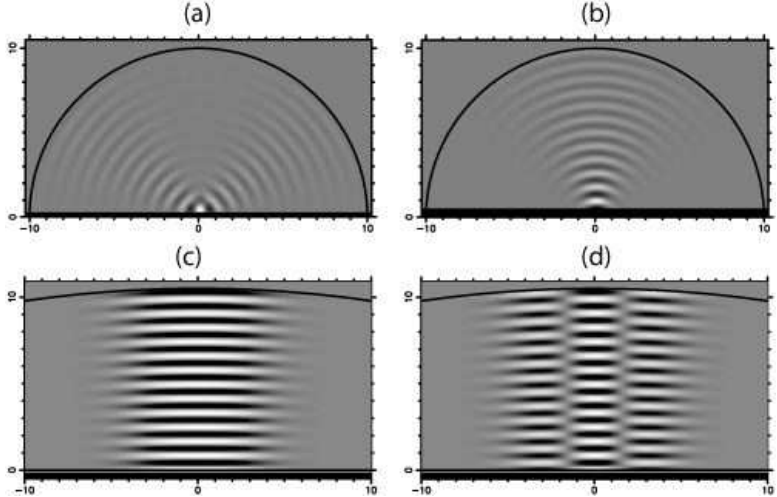}

   \end{center}
\caption{\label{cap:BraggSide}Modes with $m=1$ in the cavity bounded by
metal dome and dielectric Bragg stack. The cross sections in (a) and
(b) were obtained with $R=10\,\mu\mathrm{m}$ for modes at dome heights
(a) $h=9.782\,\mu\mathrm{m}$ and (b) $h=9.544\,\mu\mathrm{m}$. The
calculated wavelengths and $Q$-factors are (a) $\lambda=812.394$nm,
and (b) $\lambda=796.328$nm. A striking difference between (a) and
the comparable metallic-mirror mode in Fig. \ref{cap:AllMetal} (a)
is the V-shaped spatial pattern; this disappears at the slightly more
paraxial $h$ in (b). (c) and (d) show Gauss-Hermite like modes for
the same cavity shape as in Fig. \ref{cap:AllMetal} (c), (d). The modes
form a doublet at (c) $\lambda=811.063$nm and (d) $\lambda=811.068$nm.}

\end{figure}

With these modifications,
the solution algorithm is the same as described for the all-metal
cavity in Section \ref{sec:All-metal-cavity}. An important additional
piece of information that enters for the Bragg-bounded dome is the
$Q$-factor of the modes: if the Bragg reflectivities $r_{s,p}(\theta)$
are not of modulus one, there is leakage loss which makes the modes
metastable; to find these modes, their wavenumbers $k$ are now allowed
to be complex \cite{MEPchapter}. In the following we consider only
modes calculated to have $Q>10^{4}$, but omit a more detailed discussion
of the $Q$-factors because that will require addressing the competing
absorption losses in the curved metal mirror as well. 

\begin{figure}[!htb]
   \begin{center}
\includegraphics[%
  width=0.75\columnwidth]{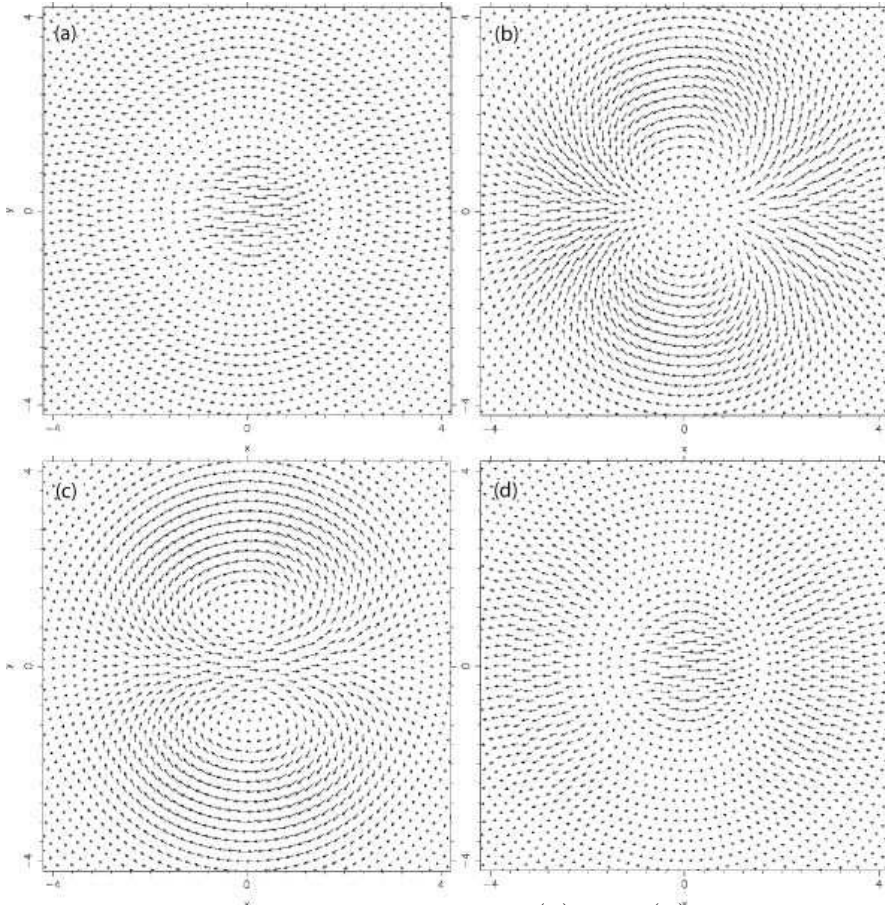}
   \end{center}
\vspace*{-.9cm}
\caption{\label{cap:TransverseBragg}Transverse polarization cross sections.
(a) and (b) correspond to the modes in Fig. \ref{cap:AllMetal} (c),
(d); the section is taken at height $z=0.25\,\mu\mathrm{m}$ and is
cropped to contain only the regions of significant field strength.
To see that this represents circular polarization, imagine the pattern
rotating rigidly, and track the resulting arrow direction at any fixed
point. Similarly, (c) and (d) correspond to the modes in Fig. \ref{cap:BraggSide}
(c), (d), but plotted at $z=0.05\,\mu\mathrm{m}$ to intersect the
domains of strongest field. }

\end{figure}
Figure \ref{cap:BraggSide} lists a set of results analogous to the
ones shown in Fig. \ref{cap:AllMetal}, but now including the effect
of the Bragg mirror. There is no significant distinction between the
two systems from a purely ray-optic point of view, and the modes in
parts (a,b) of Figures \ref{cap:AllMetal} and \ref{cap:BraggSide}
were obtained in the same way, by adiabatically following a fundamental
Gaussian mode from small to large $h$. Nevertheless, the near-hemispherical
mode of Fig. \ref{cap:BraggSide} (a) exhibits a V-shaped distribution
in $E_{y}$ which is absent for all-metal mirrors. We have found the
same behavior in cavities as large as $R=40\,\mu\mathrm{m}$, $h=39.7\,\mu\mathrm{m}$.
The mode is predominantly s-polarized, and its formation can be interpreted
as a consequence of the fact that the Bragg reflectivity $r_{s}(\theta)$
has a phase that depends on angle of incidence; in this sense, this
phenomenon is induced by the Bragg stack, but arises essentially as
a scalar wave effect \cite{David}. 

The vector nature of the electromagnetic field enters in a more intricate
way if we compare the modes shown in Fig. \ref{cap:BraggSide} (c),
(d) to the corresponding ones in Fig. \ref{cap:AllMetal} (c), (d).
For this cavity shape, all modes shown are well approximated by Eq.
(\ref{eq:veclaguerregauss}) with $m=1$ and $N=2$. However, whereas
$A=0$ or $B=0$ in Fig. \ref{cap:AllMetal}, we now find $A,B\ne0$
in both members of the doublet. Since the spatial dependence of $LG_{2}^{0}$
and $LG_{2}^{2}$ is different, polarization and orbital part of the
wave are entangled in the sense that it is impossible to factor out
a uniform polarization vector. 

The difference between the side views in Fig. \ref{cap:BraggSide} (c), (d) 
and Fig. \ref{cap:AllMetal} (c), (d) is due to the fact that the
spatial mode profiles in the dome with Bragg stack more closely resemble 
Hermite-Gauss beams\cite{Beijersbergen}, $u^{HG}_{\mu,\nu}$. This 
can be reconciled with Eq. (\ref{eq:veclaguerregauss}) by noting that
both $LG_2^2$and $LG_2^0$ have projections onto $u^{HG}_{0,2}$ and
$u^{HG}_{2,0}$. 

However, a transverse cross section of the mode field reveals the 
entanglement of polarization and spatial structure in these modes,
which is not obvious from the side views alone. In
Fig. \ref{cap:TransverseBragg}, we compare the paraxial modes from 
Figs. \ref{cap:BraggSide} and \ref{cap:AllMetal} with respect to 
the instantaneous projections of the mode field into a horizontal
plane.  Because Eq. (\ref{eq:veclaguerregauss}) is approximately
valid, one can in fact construct \cite{David} both (c) and (d) by 
appropriately superimposing the circularly polarized fields (a) and
(b). Neglecting the $z$ components of the electric field, the
resulting modes in Fig. \ref{cap:TransverseBragg} (c) and
(d) are close to circular polarization near the $z$ axis, but
generally exhibit varying polarization over the cross section. The 
vortex pair bracketing a central high-field region in 
Fig. \ref{cap:TransverseBragg} (c) corresponds to a ring of radial
(linear) polarization when the time dependence of the wave is taken
into account. 

It is remarkable in the Bragg-mirror dome that the
particular transverse polarization patterns in 
Fig. \ref{cap:TransverseBragg} (c), (d) are in fact robust over a
large range of cavity shapes $h$ and also quite insensitive to the
spectral location of the mode within the stop band of the Bragg
mirror. 

\begin{figure}[!htb]
   \begin{center}
 
\includegraphics[%
  width=0.9\columnwidth]{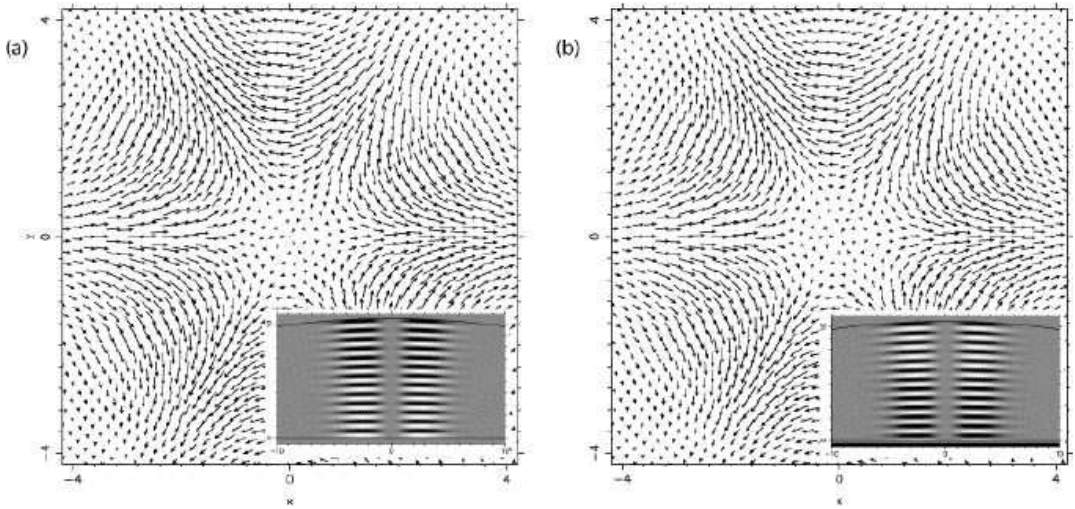}

   \end{center}
\caption{\label{cap:Transversem3}Transverse polarization cross
 sections at height $z=0.25\,\mu\mathrm{m}$ for a mode corresponding 
to $m=3$, $N=2$. This is a
 circularly polarized singlet state because $N=\vert m\vert
 -1$. Consequently, the polarization patterns with metal mirror (a) at
 $\lambda=796.091$nm  and
 Bragg stack (b) at $\lambda=811.086$ are practically indistinuishable. Insets show the
 corresponding profile of $E_x$ in the $yz$ plane. 
}

\end{figure}
The Laguerre-Gauss modes responsible for the persistent mixing with 
polarization structure given by Fig. \ref{cap:TransverseBragg} (c),
(d) have order $N=2$. For completeness, we verify that the other
remaining mode of order $2$, which has $m=3$, does not undergo the
same qualitative transition when changing from metal to Bragg
mirror. The reason is that in this case $N=\vert m\vert-1$ and thus
there is no doublet from which superpositions can be formed. The
circularly polarized singlet modes of the paraxial dome remain
circularly polarized independently of the nature of the planar
mirror. This is shown in Fig. \ref{cap:Transversem3}. 

\section{Conclusion}
There exists a great wealth of other stable but non-paraxial modes in the
dome cavity with $h<R$, which can be classified with the help of
ray-based methods as exemplified by Fig. \ref{cap:RadialMode}. 
The exact numerical calculations demonstrated in the present work are
relevant to such quasiclassical studies because we are able to treat
realistic cavities whose size is large in relation to $\lambda$. 
Future work in this direction will in particular have to address
complex boundary conditions as they arise when different types of
mirrors make up the 3D cavity. In our case, a combination of conducting
surface and dielectric multilayers is considered.

The vectorial nature of the cavity fields is essential in this
system. Fully vectorial mode calculations of the type performed here 
indicate that one of the effects induced by a Bragg-stack is the
persistent mixing of doublets illustrated in 
Fig. \ref{cap:TransverseBragg}. The nonuniform polarization patterns
can be further analyzed, in particular regarding the locations of
their singularities: in the near-paraxial situation, we pointed out
the occurence of linear polarization on rings surrounding the
circularly polarized beam axis. These considerations are of
significance in particular when the coupling between cavity field and
dipole emitters at the base of the dome is considered \cite{Jens}. 

Although we presented only results for dome mirrors in the form of a
conducting spherical shell, other shapes such as paraboloids can be
treated. By making contact with various limiting cases in this paper, the
general numerical techniques have been validated. A more detailed
discussion of the numerical methods and their implementation can be
found in a forthcoming publication\cite{David}. 

\section*{ACKNOWLEDGMENTS}
 
This work was supported by NSF Grant ECS-02-39332.



\begin{thebibliography}{10}

\bibitem{Matinaga}
F.~M. Matinaga, A.~Karlsson, S.~Machida, Y.~Yamamoto, T.~Suzuki, Y.~Kadota, and
  M.~lkeda, ``Low-threshold operation of emispherical microcavity
  single-quantum-well lasers at 4 k,'' {\em Appl. Phys. Lett.} {\bf 62},
  pp.~443--445, 1993.

\bibitem{Meissner}
M.~Aziz, J.~Pfeiffer, and P.~Meissner, ``Modal behaviour of passive, stable
  microcavities,'' {\em Phys. Stat. Sol. (a)} {\bf 188}, pp.~979--982, 2001.

\bibitem{milburn}
T.~M.Stace, G.~J. Milburn, and C.~H.~W. Barnes, ``Entangled two-photon source
  using biexciton emission of an asymmetric quantum dot in a cavity,'' {\em
  Phys. Rev. B} {\bf 67}, p.~085317, 2003.

\bibitem{Jens}
J.~U. N{\"o}ckel, G.~Bourdon, E.~L. Ru, R.~Adams, J.-M.~M. I.~Robert, and
  I.~Abram, ``Mode structure and ray dynamics of a parabolic dome
  microcavity,'' {\em Phys. Rev. E} {\bf 62}, pp.~8677--8699, 2000.

\bibitem{bulovic}
V.~Bulovic, V.~G. Kozlov, V.~B. Khalfin, and S.~R. Forrest,
  ``Transform-limited, narrow-linewidth lasing action in organic semiconductor
  microcavities,'' {\em Science} {\bf 279}, pp.~553--555, 1998.

\bibitem{coyle}
S.~Coyle, G.~V. Prakash, J.~J. Baumberg, M.~Abdelsalem, and P.~N. Bartlett,
  ``Spherical micromirrors from templated self-assembly: Polarization rotation
  on the micron scale,'' {\em Appl. Phys. Lett.} {\bf 83}, pp.~767--769, 2003.

\bibitem{Dorn}
R.~Dorn, S.~Quabis, and G.~Leuchs, ``Smaller, sharper focus for a radially
  polarized light beam,'' {\em Phys. Rev. Lett.} {\bf 91}, p.~233901, 2003.

\bibitem{thesis}
J.~U. N{\"o}ckel, {\em PhD thesis}, Yale University, 1997.

\bibitem{David}
D.~H. Foster and J.~U. N{\"o}ckel, ``Methods for 3-d vector microcavity
  problems involving a planar dielectric mirror,'' {\em Opt. Commun.} {\bf
  234}, pp.~351--383, 2004.

\bibitem{mekis}
A.~Mekis, J.~U. N{\"o}ckel, G.~Chen, A.~D. Stone, and R.~K. Chang, ``Ray chaos
  and q-spoiling in lasing droplets,'' {\em Phys. Rev. Lett.} {\bf 75},
  pp.~2682--2685, 1995.

\bibitem{lacey}
S.~Lacey, H.~Wang, D.~Foster, and J.~N{\"o}ckel, ``Directional tunnel escape
  from nearly spherical optical resonators,'' {\em Phys. Rev. Lett.} {\bf 91},
  p.~033902, 2003.

\bibitem{Siegman}
A.~E. Siegman, {\em Lasers}, University Science Books, Sausalito, CA, 1986.

\bibitem{Beijersbergen}
M.~W. Beijersbergen, L.~Allen, H.~E. L.~O. van~der Veen, and J.~P. Woerdman,
  ``Astigmatic laser mode converters and transfer of orbital angular
  momentum,'' {\em Opt. Commun.} {\bf 96}, pp.~123--132, 1993.

\bibitem{Yeh}
P.~Yeh, {\em Optical waves in layered media}, Wiley, New York, 1988.

\bibitem{MEPchapter}
J.~U. N{\"o}ckel and R.~K. Chang, ``2-d microcavities: Theory and
  experiments,'' in {\em Cavity-Enhanced Spectroscopies},  R.~D. van Zee and
  J.~P. Looney, eds., {\em Experimental Methods in the Physical Sciences} {\bf
  40}, pp.~185--226, Academic Press, San Diego, 2002.

\end{thebibliography}
\end{document}